\documentclass[%
reprint,
 amsmath,amssymb,
 aps,
]{revtex4-1}

\usepackage[colorlinks=true,breaklinks=true]{hyperref}
\usepackage[normalem]{ulem}
\usepackage[utf8]{inputenc}
\hypersetup{allcolors=[rgb]{0.0 0.0 0.6},linkcolor=[rgb]{0.75 0.05 0.05}}
\usepackage{amsmath,amssymb}
\usepackage{epsfig}  
\usepackage{graphicx}   
\usepackage{slashed}             
\usepackage{url}
\usepackage{color}
\usepackage{multirow}
\usepackage[dvipsnames]{xcolor}
\usepackage{letltxmacro}

\usepackage{graphicx}
\usepackage{dcolumn}
\usepackage{bm}
\usepackage[mathlines]{lineno}

\newcommand{\apjs}{Astrophys. J. Suppl. Ser.}
\newcommand{\apjl}{Astrophys. J. Lett.}
\newcommand{\mnras}{Mon. Not. R. Astron. Soc.}
\newcommand{\aap}{Astron. Astrophys.}

\begin{document}

\preprint{APS/123-QED}

\title{Spontaneous Scalarization as a New Core-Collapse Supernova Mechanism\\ and its Multi-Messenger Signals}

\author{Takami Kuroda}
 \email{takami.kuroda@aei.mpg.de}
\author{Masaru Shibata}
 \email{mshibata@aei.mpg.de}
\affiliation{%
 Max-Planck-Institut f{\"u}r Gravitationsphysik, Am M{\"u}hlenberg 1, D-14476 Potsdam-Golm, Germany
}%

\date{\today}

\begin{abstract}
We perform multi-dimensional core-collapse supernova (CCSN) simulations in a massive scalar-tensor theory for the first time with a realistic equation of state and multi-energy neutrino radiation. 
Among the set of our models varying the scalar mass and the coupling strength between the scalar and gravitational fields, a particular model allows for recurrent spontaneous scalarizations (SSs) in the proto-neutron star (PNS).
Each SS induces the PNS collapse and subsequent bounce, from which devastating shock waves emanate and eject the PNS envelope.
The explosion energy can easily exceed $\mathcal O(10^{51})$\,erg.
This study reveals new aspects of SS as the explosion mechanism of CCSNe.
We also discuss its characteristic multi-messenger signals: neutrinos and gravitational waves.
\end{abstract}

\maketitle


\section{Introduction}
\label{sec:Introduction}
General relativity (GR) is currently the standard theory of gravity.
It has successfully explained a large number of precision tests to date \citep{GR_Test_Will14,GR_Test_Psaltis08,GR_Test_berti15}.
However, the discovery of inflation and the fact that most of the energy content of the universe takes the form of dark energy and dark matter \citep{Perlmutter99,Komatsu7yrs,Abbott19,PlanckCollaboration20} have led us to consider alternatives to GR.

One of the simplest and cosmologically and astrophysically motivated alternatives is the scalar-tensor (ST) theories of gravity, proposed in the seminal works of 
\citep{Jordan59,Fierz56,Brans61}.
In the ST theories, an additional scalar sector is added to the field equations, preserving the consistency with GR in the weak-field regime, while significant deviations are allowed in the strong-field regime.
A major example is spontaneous scalarization (SS) in neutron stars (NSs) \citep{Damour93,Ramazanoglu16,Morisaki&Suyama17}.
The SS occurs by a non-linear coupling between the scalar and gravitational fields, which enables exponential amplification of the scalar field, and changes the gravitational field as well as the NS structure \citep{Doneva18}.
Previous studies considering the hydrostatic cold NSs reported that the scalarization might significantly modify the mass-radius relation of NSs from that in GR \citep{Ramazanoglu16,Morisaki&Suyama17,Sotani17,Staykov18,Rosca-Mead20}. 

The SS may also take place in the PNS formed in the aftermath of massive stellar core-collapse (CC).
Refs. \cite{Novak00,Gerosa16,Sperhake17,Cheong19,Rosca-Mead20} conducted CC simulations in the ST theories, focusing mainly on the scalar-type gravitational wave (GW) emissions.
After the PNS formation, it starts contraction due to continuous mass accretions and increases its density.
Eventually, the coupling between the scalar and gravitational fields enters the non-linear phase and facilitates the exponential growth of the scalar field.
Depending on how the scalar field couples with the gravitational field, the PNS sometimes changes its structure not steadily but dynamically similar to the CC, and liberates a significant amount of gravitational potential energy.
This might be a remarkable feature in terms of CCSN dynamics, as the second collapse and bounce may produce strong shock waves \citep{Rosca-Mead20}.
Interestingly, the SS may happen multiple times and leave behind various compact stars \citep{Rosca-Mead20}.
The amplified scalar-type GWs propagate outward from the PNS core and could be reached us with sizeable amplitudes \citep{Sperhake17,Cheong19}.

To date, however, all the CCSN simulations of massive stellar collapse in the ST theories were performed in spherical symmetry with very simplified EOSs \cite{Novak00,Gerosa16,Sperhake17,Cheong19,Rosca-Mead20}, although it is well-known that the multi-dimensionality is the key for the successful CCSN explosion.
Moreover, the neutrino radiation, which is most crucial for the PNS evolution as well as for the CCSN dynamics, was completely neglected.
It is also not well understood what the potential impacts of the SS on the explosion dynamics are.

In this study, we conduct axisymmetric CCSN simulations of a massive star in a {\it massive} scalar-tensor (MST) theory for the first time with a realistic EOS and multi-energy neutrino radiation.
The main motivation for considering the massive scalars is that most of the parameters in the ST theories, which describe how strong the scalar and gravitational fields couple, are strongly constrained in the massless case through binary pulsar observations \citep{Freire12,Antoniadis13,Zhao22}, while the constraint can be significantly loosened in the presence of a massive scalar~\citep{Ramazanoglu16,Morisaki&Suyama17}.

This paper is organized as follows.
Section~\ref{sec:Formalism} starts with a concise summary of our GR radiation-hydrodynamic (RHD) scheme in MST theory and also describe the initial setup of the simulation.
The main results and detailed analysis of the effects of phase transition are presented in Section~\ref{sec:Results}.
We summarize our results and conclude in Section~\ref{sec:Conclusions and Discussions}.
Geometric units $c=G=1$ are used in Sec.~\ref{sec:Formalism} section, while cgs units are used in the rest.
Greek indices run from 0 to 3 and Latin indices from 1 to 3, except $\nu$ and $\varepsilon$ which denote neutrino species and energy, respectively.

\section{Formalism}
\label{sec:Formalism}
We perform CCSN simulations with a full relativistic multi-energy neutrino transport in the Jordan frame \citep{Jordan59,Brans61}.
In this frame, all the basic equations are derived from variation of the action $\mathcal S$ expressed as (using geometrical units $G=c=1$)
\begin{eqnarray}
    \mathcal S&=&\frac{1}{16\pi}\int \sqrt{-g} d^4x\left[
    \phi R-\frac{\omega(\phi)}{\phi}
    g^{\alpha\beta}\nabla_\alpha \phi \nabla_\beta \phi\right. \nonumber \\
    && \left. -\frac{4m^2}{B \hbar^2}\phi^2 \ln \phi
    \right]+\mathcal{S}_{\rm \nu m},
\label{eq:action}
\end{eqnarray}
where $g$, $R$, and $\nabla_\alpha$ denote the determinant, Ricci scalar, and covariant derivative associated with the spacetime metric $g_{\alpha\beta}$; $\phi$ is a real scalar field; $\omega(\phi)$ determines the strength of the coupling between the gravitational and scalar fields; $m$ is the scalar field mass; $B$ is a dimensionless free parameter; $\mathcal{S}_{\rm \nu m}$ is the contribution from the neutrino radiation and matter fields.
Regarding $\omega(\phi)$, we adopt the form of \cite{Damour93,Shibata_JFBD_14}
\begin{equation}
    \frac{1}{\omega(\phi)+3/2}=B \ln \phi \label{eq:omega_B}.
\end{equation}

Following \cite{Shibata_JFBD_14}, new scalar fields $\varphi$, redefined from $\phi\equiv\exp(\varphi^2/2)$, and $\Phi\equiv-n^\alpha\nabla_{\alpha}\varphi$ are evolved as:
\begin{eqnarray}
        \left(\partial_t\right.&-&\left.\beta^i\partial_i\right)\varphi=-\alpha \Phi,\label{eq:varphi}\\
    \left(\partial_t\right.&-&\left.\beta^i\partial_i\right)\Phi=-\alpha D^i D_i \varphi-(D_i \alpha) D^i\varphi+\alpha K \Phi \nonumber\\
    &-&\alpha\varphi\left(D_i\varphi D^i\varphi -\Phi^2\right)+2\pi\alpha BT\varphi \exp(-\varphi^2/2)\nonumber \\
    &+&\alpha (m/\hbar)^2 \varphi \exp(\varphi^2/2)\label{eq:Phi}.
\end{eqnarray}
Here, $\alpha$ is the lapse function; $\beta^i$ is the shift vector; $D^i$ is the covariant derivative with respect to the 3-metric $\gamma_{ij}$; $K$ is the trace of the extrinsic curvature;  and $T\equiv g_{\alpha\beta}T^{\alpha\beta}$, where $T^{\alpha\beta}$ denotes the total stress-energy tensor considering the matter $T_{\rm m}^{\alpha\beta}$ and neutrino radiation field $T_{(\nu,\varepsilon)}^{\alpha\beta}$ \citep{KurodaT16,KurodaT22}:
\begin{equation} T^{\alpha\beta} = 
T_{\rm m}^{\alpha\beta} +\int d\varepsilon \sum_{\nu\in\nu_e,\bar\nu_e,\nu_x}T_{(\nu,\varepsilon)}^{\alpha\beta}.
\label{TotalSETensor}
\end{equation}

In our CCSN simulations, we employ the Z4c formalism \citep{Hilditch13}, when the system is in GR, i.e. $\varphi\ll1$, to propagate away local violaiton of the Hamiltonian constraint $\mathcal H$ written by
\begin{eqnarray}
    \mathcal {H}&=&R-\tilde A^{ij}\tilde A_{ij}+\frac{2}{3}(\hat K+2\Theta)^2-16\pi\phi^{-1}\rho_{\rm H}\nonumber \\
    &&-\frac{4m^2}{B\hbar^2}\phi\ln \phi
    -\omega\phi^{-2}\left[\Pi^2+(D_i\phi)D^i \phi \right]\nonumber \\
    &&-2\phi^{-1}\left(-K\Pi+D_i D^i\phi \right), 
    \label{eq:Hamiltonian}
\end{eqnarray}
where $\hat K\equiv K-2\Theta$ and $\Pi\equiv-n^\alpha\nabla_\alpha\phi$, with $\Theta$ and $n^\alpha$ being an auxiliary variable \citep{Hilditch13} and the unit normal to spatial hypersurfaces, respectively.
$\tilde A_{ij}$ is the trace free part of the conformal extrinsic curvature and $\rho_{\rm H}\equiv n_\alpha n_\beta T^{\alpha\beta}$.
Once the system deviates from GR as $|\varphi|\gtrsim\mathcal O(0.1)$, we enforce $\Theta=\partial_t \Theta=0$, which recovers the BSSN formalism.
Afterward, we monitor the growth of $\mathcal H$.

\textit{Models and parameters.---}We perform axisymmetric CC simulations to a non-rotating $50\,{\rm M}_\odot$ progenitor star of \citep{Umeda08}.
It was used in the previous studies of \citep{Fischer18,KurodaT22} to explore the impacts of a first-order QCD phase transition.
We use the DD2 EOS of \cite{Typel10}.

We have two free parameters, $B$ and $m$, which are chosen to satisfy the current observational constraints.
$B$ has a relation to $\beta(\beta_0)$, which is frequently used in other literatures \citep{Damour93,Ramazanoglu16,Cheong19,Rosca-Mead20}, as $B=-2\beta$ \citep{Shibata_JFBD_14}.
According to \citep{Ramazanoglu16}, $\beta$ is weakly bounded to avoid the scalarization in white dwarfs (WDs), while allowing for it in NSs, which can be translated into
\begin{equation}
    6\lesssim B\lesssim \mathcal O(10^3) \label{eq:B}.
\end{equation}
The scalar mass is also constrained within
\begin{equation}
10^{-16}\,{\rm eV}\lesssim m \lesssim 10^{-9}\,{\rm eV} \label{eq:m},
\end{equation}
where the lower band comes from cosmological effects and binary observations \citep{Freire12,Antoniadis13} and the upper one from the assumption that the NS can be scalarized \citep{Ramazanoglu16}.
\citep{GRTESTS_GW170817} reports that GW170817 offers rather weaker constrains on ST theories than, e.g., binary pulsar PSR J0737-3039 does \citep{Freire12,Antoniadis13,Zhao22}, making above parameter ranges also consistent with GW170817.
Regarding the initial (or asymptotic) value of $\varphi_0$, we assume a uniform weak field $\varphi(t=0)=\varphi_0=10^{-14}$. 
With this choice, the emission of scalar-type GWs is significantly suppressed (see below).

We present results of CC simulations for three models with $(B,m\,[\rm{eV}])=(20,10^{-11})$, $(10,10^{-14})$, and $(20,10^{-14})$; hereafter $B20m11$, $B10m14$, and $B20m14$, respectively.
Furthermore, as a reference GR model with the same $50\,{\rm M}_\odot$ progenitor star and a DD2-based QCD EOS \citep{Bastian:2021}, a model $QCD(GR)$ \citep[cf.][]{KurodaT22} is also introduced.
We mention that these parameters predict significant deviations from GR in the context of binary NS simulations with {\em massless} scalar fields
\citep{Shibata_JFBD_14,Barausse13}.

\section{Results}
\label{sec:Results}
\begin{figure}[t!]
\begin{center}
\includegraphics[angle=-90.,width=\columnwidth]{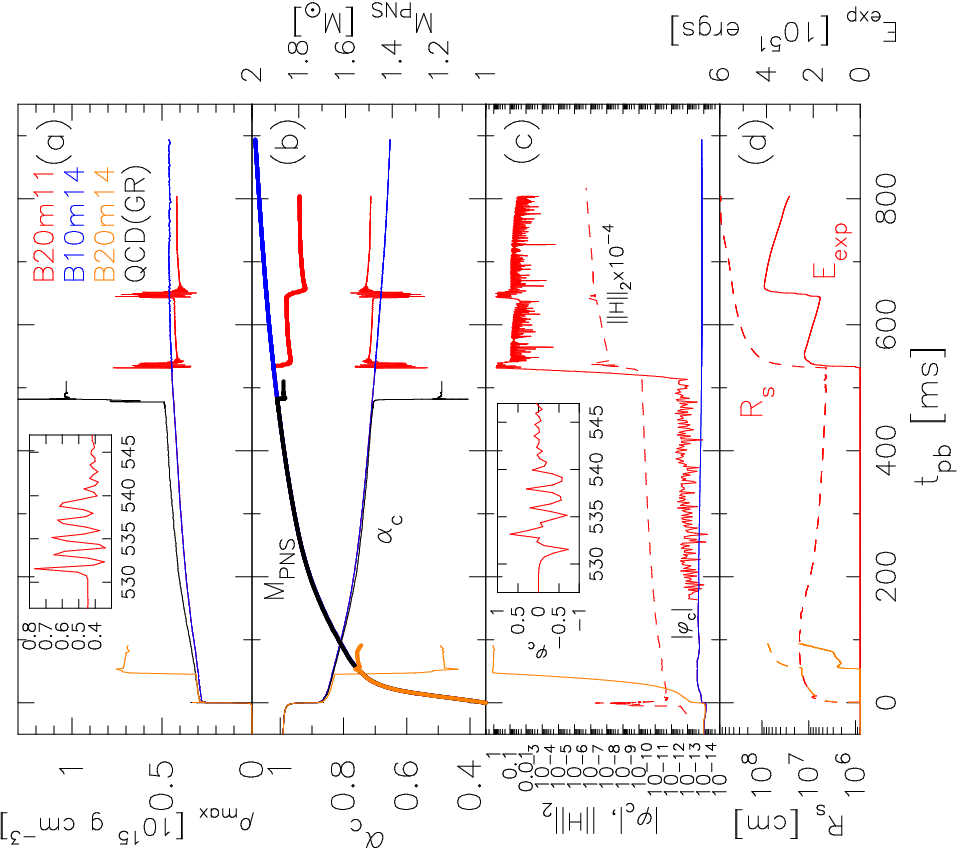} 
\caption{Overall evolution feature of all models. Panel (a): the maximum rest-mass density $\rho_{\rm max}$; (b): PNS rest mass $M_{\rm PNS}$ and central lapse function $\alpha_{\rm c}$; (c): central scalar field $\varphi_{\rm c}$ and 2-norm of the Hamiltonian constraint $||\mathcal{H}||_2\times10^{-4}$; (d): averaged shock radius $R_{\rm s}$ (dashed) and diagnostic explosion energy $E_{\rm exp}$ (solid).
The color represents each model listed in panel (a).
\label{fig:Overall}}
\end{center}
\end{figure}

We begin with a description of overall dynamics.
Panels (a)--(d) in Fig.~\ref{fig:Overall} depict: (a) the maximum rest-mass density $\rho_{\rm max}$; (b) the PNS baryon rest mass $M_{\rm PNS}$ (thick lines) and central lapse function $\alpha_{\rm c}$ (thin); (c) the central scalar field $\varphi_{\rm c}$ and 2-norm of the Hamiltonian constraint $||\mathcal{H}||_2$ divided by $10^4$; and (d) averaged shock radius $R_{\rm s}$ and diagnostic explosion energy $E_{\rm exp}$.
Insets in (a) and (c) show a magnified view at the first SS.
Note that in (c) and (d) we plot the results only for representative models.

In $B20m11$ and $B10m14$, $\varphi_{\rm c}$ essentially keeps its initial value $|\varphi_{\rm c}|\sim10^{-14}$ till $t_{\rm pb}\sim500$\,ms~\footnote{We restart the model $B20m11$ using the data of $B10m14$ at $t_{\rm pb}\sim150$\,ms, that is why $\varphi_c$ in $B20m11$ shows an identical evolution with that in $B10m14$ at $t_{\rm pb}\lesssim150$\,ms.}, where $t_{\rm pb}$ measures the post (first-)bounce time.
This indicates that they are essentially in GR and thus show a similar evolution to $QCD(GR)$.
At $t_{\rm pb}\sim520$\,ms, $\varphi_{\rm c}$ in $B20m11$ exponentially grows from $\mathcal O(10^{-14})$ to $\mathcal O(1)$ within a few ms.
This is the moment of the first SS.
Following the analysis in the massless case (Sec.~III.A. of~\cite{Shibata_JFBD_14}), the SS in the presence of scalar mass occurs when $k^2>0$ and $kR\rightarrow \pi/2$ are satisfied, where $k^2\equiv-\left[2\pi BT+(m/\hbar)^2\right]$ and $R$ denotes the PNS radius.
Along with the PNS contraction, $-T\sim \rho$ increases, $|2\pi BT|$ exceeds $(m/\hbar)^2$, and eventually $kR$ approaches $\pi/2$, which induces the SS.
In the previous literature, e.g., \cite{Rosca-Mead20}, if the SS does not happen at the first core bounce, i.e., $|\varphi_c|\ll1$ at $t_{\rm pb}=0$, $\varphi_c$ shows essentially no increment afterward.
We attribute this due to the absence of PNS contraction with simplified microphysics.
In $B20m14$, the light scalar mass of $10^{-14}$\,eV allows for an even faster SS than $B20m11$ already at $t_{\rm pb}\sim50$\,ms.
On the other hand in $B10m14$, the SS is not observed during our simulation time.

When the SS happens, $\phi$ deviates from unity, which prompts the second collapse and bounce.
The second-bounce produces strong shock waves (panel (d)) and unbound some of the PNS materials amounting to $\sim0.1$\,$M_\odot$ (panel (b)).
Panel (d) shows that those ejecta possess $E_{\rm exp}\sim2\times10^{51}$\,ergs at $t_{\rm pb}\sim520$\,ms in $B20m11$.

After the first SS, we find an interesting phenomenon in $B20m14$: recurrent SSs.
At $t_{\rm pb}\sim630$\,ms, $\varphi_c$ suddenly increases again, which induces the (third) collapse.
The reason of the second SS can be explained by the decrease of $M_{\rm PNS}$ at the first SS and also by the value of $B=20$.
As discussed in \cite{Shibata_JFBD_14}, the requisites for the SS inside NSs are a sufficiently high compactness of PNS and a large value of $-T\sim \rho$.
Therefore, the decreases both in compactness and density hinder the scalar waves condensation.
Indeed from panel (c), $\varphi_c$ quickly decreases from $\mathcal O(1)$ to $\lesssim\mathcal O(10^{-1})$ after the first SS.
Afterward, however, the mass accretion still continues and increases both $\rho_{\rm c}$ and $M_{\rm PNS}$, resulting in the second SS at $t_{\rm pb}\sim630$\,ms.
The second SS again induces strong shock waves and boosts $E_{\rm exp}$ by $\sim2\times10^{51}$\,erg.
Also in $B20m14$, we witness the explosion.
However, the prompt SS, which occurs before the shock stagnation and thus during when the mass accretion rate is still high, avoids the significant PNS mass loss and $\varphi_c$ keeps $\mathcal O(1)$ afterward.

Regarding the Hamiltonian constraint violation, $||\mathcal H||_2$ is kept at $\mathcal O(10^{-7})$ before the first SS with the help of the Z4c formalism.
Even after we switch the calculation to the BSSN method, it stays at $\lesssim\mathcal O(10^{-2})$ without any drastic increase.

\textit{Multi messenger signals.---}Next we discuss multi-messenger signals.
\begin{figure}[t!]
\begin{center}
\includegraphics[angle=-90.,width=\columnwidth]{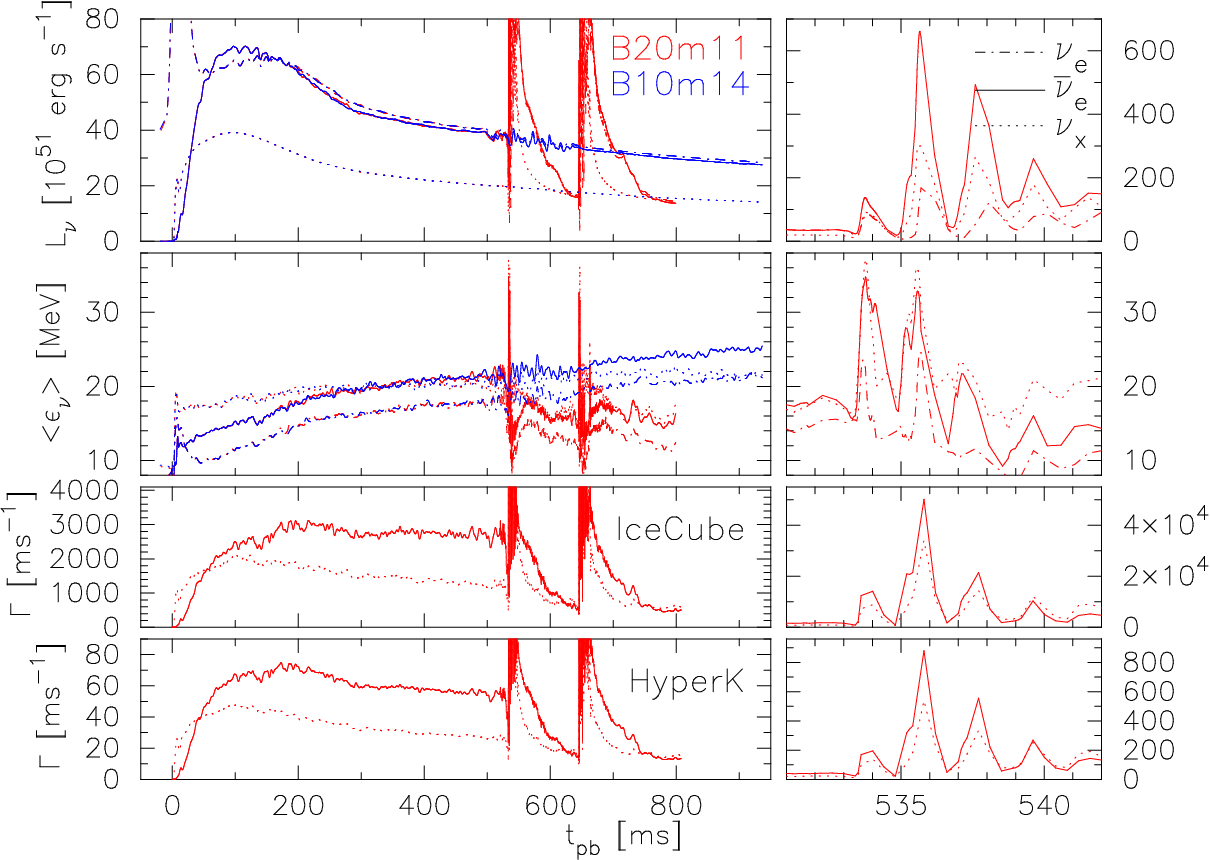} 
\caption{Various neutrino profiles as functions of $t_{\rm pb}$. The right column shows a magnified view at the first SS for $B20m11$. From top: the neutrino luminosity $L_\nu$; mean energy $\langle\epsilon_\nu\rangle$; and neutrino detection rate $\Gamma$ of IC and HK, respectively.
All plots assume a source distance of $D=10$\,kpc.
The color and line style represents the model and neutrino species, respectively, shown in the top panels.
\label{fig:LnuEnu}}
\end{center}
\end{figure}
We begin with the neutrino signals for representative models $B20m11$ and $B10m14$.
Fig.~\ref{fig:LnuEnu} displays from top: the neutrino luminosity $L_\nu$; mean energy $\langle\epsilon_\nu\rangle$; and neutrino detection rate $\Gamma$ of IceCube (IC) \citep{abbasi11,salathe12} and Hyper-Kamiokande (HK) \citep{abe11,HK18}.
The neutrino detection rate $\Gamma$ is evaluated in the same way as \citep{KurodaT22}.
We assume a source distance of $D=10$\,kpc.
In the right column, we plot a magnified view of the first SS in $B20m11$.

The red lines show a clear fingerprint of the SS.
At $t_{\rm pb}\sim520$ and $630$\,ms we observe neutrino bursts whose peak luminosities reach $\sim2$--$6\times10^{53}$\,erg\,s$^{-1}$ with 
$L_{\bar\nu_e}$ (solid line) and $L_{\nu_e}$ (dash-dotted) showing the highest and lowest luminosity, respectively, among the 6 species.
The hierarchy is simply due to the propagation of strong shock waves through the neutron rich environment and is analogous to that in QCD models \citep{Fischer20,KurodaT22}.
The peak count rates $\Gamma$, evaluated by $L_\nu$ and $\langle \epsilon_\nu\rangle$, reach $\sim3000$--$5000$\,ms$^{-1}$ (IC) and $\sim600$--$900$\,ms$^{-1}$ (HK).
Compared to the QCD models, multi-neutrino bursts (with more than three times) would be a clear indication of recurrent SSs, as CCSNe with QCD phase transition experience the second collapse and bounce only once \citep{Fischer18,Zha20,KurodaT22}.

\begin{figure}[t!]
\begin{center}
\includegraphics[angle=-90.,width=\columnwidth]{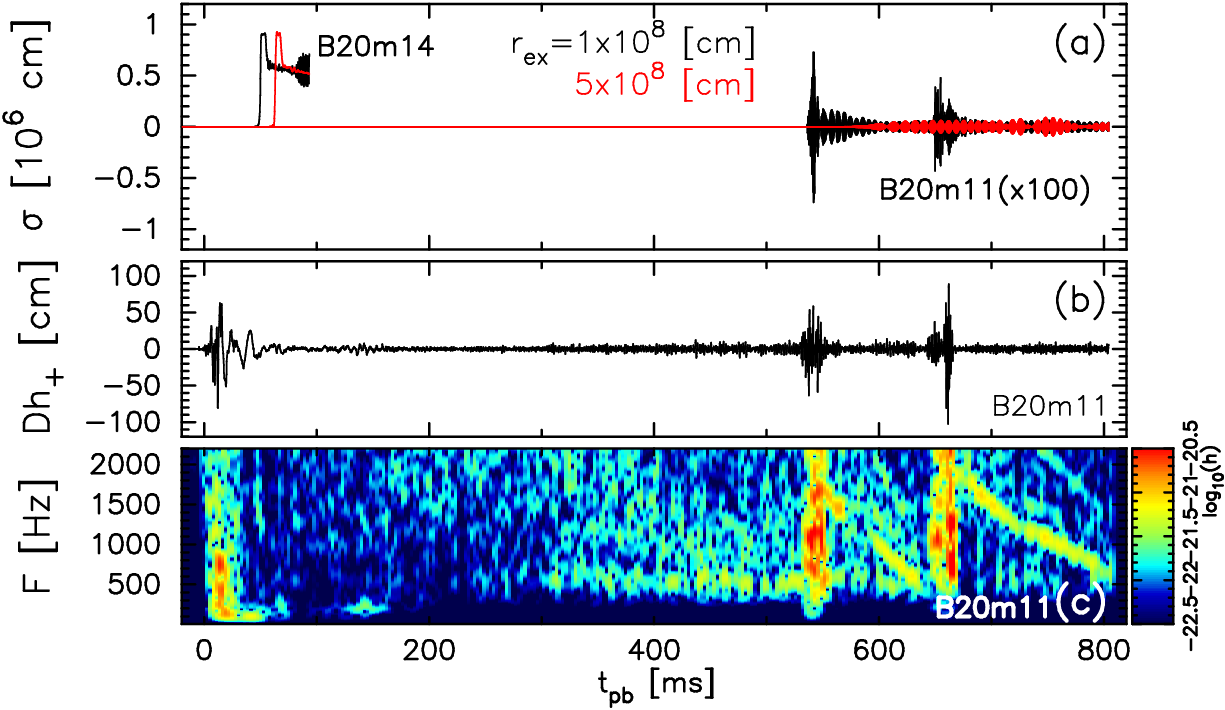}  
\caption{Panel (a): the scalar waveform $\sigma\equiv r_{\rm ex}\varphi$ extracted at two different radii $r_{\rm ex}=10^8$ (black line) and $5\times10^8$\,cm (red); (b): matter origin GWs $Dh_+$; (c): spectrogram $\tilde h(F,t)$ of $h_+$. Here we assume a source distance of $D=10$\,kpc.
Panel (a) shows $\sigma$ of $B20m14$ and $B20m11$, where $\sigma$ of $B20m11$ is multiplied by 100.
Panels (b,c) show only of $B20m11$.
\label{fig:GW}}
\end{center}
\end{figure}

Finally, we discuss scalar-wave and GW emissions.
Fig.~\ref{fig:GW} presents: (a) the scalar waveform $\sigma\equiv r_{\rm ex}\varphi$ extracted at two different radii $r_{\rm ex}(=x)=10^8$ (black line) and $5\times10^8$\,cm (red) for $B20m11$ and $B20m14$; (b) matter origin GWs $Dh_+$; and (c) spectrogram of $h_+$ assuming $D=10$\,kpc obtained by a short-time Fourier transform.
Panels (b) and (c) show the results only for $B20m11$.
Note that the {\em scalar-type} GWs, whose amplitudes can be $r_{\rm ex}(\phi-\phi_0)$ with $\phi_0 (\approx 1)$ being the asymptotic value of $\phi$ and should be distinguished from $\sigma$, are essentially suppressed in the current context, as the amplitudes in the far zone $r_{\rm ex}(\phi-\phi_0)\sim r_{\rm ex}\varphi_0(\varphi-\varphi_0)$ are quite small for the current value of $\varphi_0=10^{-14}$ \citep[for the scalar-type monopole GWs from CCSNe, see][]{Gerosa16,Sperhake17,Cheong19,Rosca-Mead20}.

Panel (a) depicts the scalar wave propagation and the influence of scalar mass on it.
In $B20m11$, remarkably large scalar wave amplitudes reaching $|\sigma|\sim5\times10^3$\,cm are observed at $r_{\rm ex}=10^8$\,cm.
However, the scalar waves subside quickly at more distant radii, e.g., $|\sigma|\sim800$\,cm and $\sim70$\,cm at $r_{\rm ex}=5\times10^8$\,cm
and $10^9$\,cm, respectively.
This is due to the presence of mass term, which creates a critical frequency $\omega_\ast\equiv m/\hbar$ below which all scalar-wave modes exponentially decay \citep{Sperhake17}.
$B20m11$ employs a relatively large scalar mass $m=10^{-11}$\,eV corresponding to $\omega_\ast \approx 1.5\times10^4$\,Hz and thus most of the relevant scalar waves are damped.
On the other hand in $B20m14$, $\varphi$ propagates obeying $\partial_r(\sigma)\sim0$ because of the low cut-off frequency $\omega_\ast \approx 1.5\times10$\,Hz.

Regarding matter origin GWs, we observe strong GW bursts with $|Dh|\sim50$--$100$\,cm.
In comparison to the QCD model, for which $|Dh|\sim250$\,cm \citep{Zha20,KurodaT22}, the wave amplitudes in $B20m11$ are a few times smaller.
This is due to the weaker core bounce in the MST models and the convection motions are weakened in $B20m11$ \citep{KurodaT22}.

\begin{figure}[t!]
\begin{center}
\includegraphics[angle=-90.,width=\columnwidth]{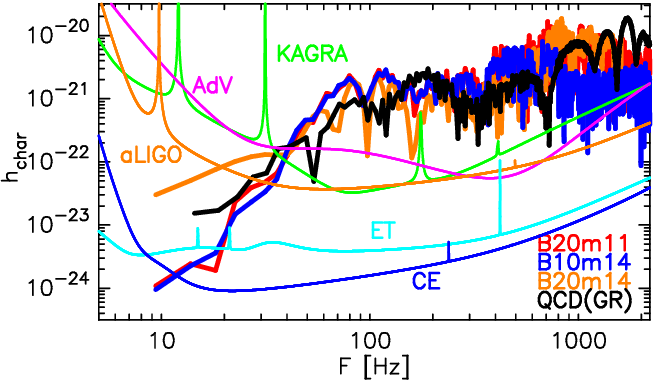}  
\caption{Characteristic strain of matter origin GWs (thick) overplotted by the sensitivity curves of the current- and third-generation GW detectors  (thin): advanced LIGO, advanced VIRGO, KAGRA, Einstein Telescope, and Cosmic Explorer. We assume a source distance of 10\,kpc.
\label{fig:Sensitivity}}
\end{center}
\end{figure}
Fig.~\ref{fig:Sensitivity} shows the detectability of matter origin GWs assuming $D=10$\,kpc.
We overplot the sensitivity curves of the current- and third-generation GW detectors: advanced LIGO (aLIGO), advanced VIRGO (AdV), KAGRA \citep{abbott18det}; Einstein Telescope (ET) \citep{ET_Hild}; and Cosmic Explorer (CE) \citep{CE_Abbott}.
Among the MST models that experience the SS(s), $B20m11$ and $B20m14$ exhibit excesses at $F\sim700$--$800$\,Hz, which are originated from the CC and bounce in association with the SS.
An approximate estimation of the signal-to-noise ratio of these models gives $\sim 100$ for the peak component at $F\sim700$\,Hz even for the current GW detectors and $\sim1000$ for the third-generation detectors, though more detailed analyses are essential \citep[e.g.][]{Hayama15,Gossan16}.
Again because of the weaker core bounce in MST models, its peak frequency $F\sim700$\,Hz is substantially lower than $\sim1.5$\,kHz in $QCD(GR)$, making it appear within the best sensitivity of GW detectors considered.
Furthermore, the duration time of strong GW emissions in $B20m11$, $20$--$30$\,ms (see Fig.~\ref{fig:GW}), which should be also multiplied by two events, is considerably longer than $\sim4$\,ms in the QCD model \citep{KurodaT22}.
These two facts, namely the lower peak frequency and longer duration time, are beneficial to the detection of these signals,  though more detailed analyses are essential \citep[e.g.][]{Hayama15,Gossan16}.

\section{Conclusions and Discussions}
\label{sec:Conclusions and Discussions}
We presented the results of the first multi-D CCSN simulations with neutrino radiation in a MST theory. Our models demonstrated dramatic impacts of the SS on the explosion dynamics as well as on the multi-messenger signals.
In $B20m11$, we observe multiple SSs, with each inducing the PNS collapse and producing strong bounce shock waves.
Those shock waves eject a part of the PNS envelope, whose explosion energy reaches $\sim2\times10^{51}$\,erg for each SS event.
In the present study, we consider a non-rotating progenitor star, which results in an essentially spherical shock expansion.
Although many of energetic SN events accompany aspherical explosion morphology \cite{Maeda08,Mazzali14}, which could possibly be explained by such as magnetorotational explosion \cite{Obergaulinger17,KurodaT20}, there exist some exceptions, e.g., SLSN I LSQ14mo \cite{SLSN_I_LSQ14mo}, for which nearly spherical explosion is reported \cite{Leloudas15}.
Such events might be explained by the present MST models.

The SSs imprint their signatures in neutrinos and GWs.
Each SS triggers a strong neutrino burst, which is originated from the same mechanism as in the QCD model \cite{Fischer18,Zha20,KurodaT22} and is observable for the Galactic events.
Although it is unlikely to detect scalar-type GWs from the current models, the matter-origin GWs in association with the SS are strong enough for the current- and third-generation detectors for galactic events.
A combination of the relatively low peak frequency and long GW emission time with possibly multiple events is beneficial to the detection of these peculiar GWs.

As a final remark, this study reports only a limited number of models using one progenitor star.
Although our scenario is sensitive to the parameter set $(B,m)$, which still has a large uncertainty as Eqs.~(\ref{eq:B}, \ref{eq:m}) indicate, the progenitor dependence may not be so strong as long as the progenitor is sufficiently massive to overcome the repulsive potential of massive scalar fields.
This is because, the SS takes place when $kR\sim\sqrt{2\pi B\rho}R\sim\sqrt{2\pi BM_{\rm PNS}/R}\rightarrow\pi/2$ and the compactness parameter $M_{\rm PNS}/R$ varies only a few 10\,\% among various progenitor models \cite{Ott18,KurodaT17}.
Therefore the primal factor to the SS is the $B$-parameter.
On the other hand for less massive progenitor stars, we do not anticipate that the current SS models would greatly modify their canonical explosion scenario, namely the neutrino heating explosion \cite{Janka16,BMulelr20_review}.
This can be supported from that the less massive PNS with $\sim1.4$\,M$_\odot$ experiences only a weak scalarization \cite{Ramazanoglu16,Rosca-Mead20} and also from that our previous SN simulations for less massive progenitor stars explode by neutrino heating shortly after the first core bounce, suppressing their PNS mass growth at $\sim1.4$\,M$_\odot$ \cite{KurodaT22}.

In this study, to explore possible SN explosion scenarios in MST theory within a reasonable simulation time, the employed parameters might deviate significantly from yet-to-be-known actual values, though they are still within the current constraint.
In addition, as we have mentioned, many of observed SNe with energetic explosion energy are accompanied by aspherical explosion morphology, indicating that the presented SN scenario in MST theory are likely not to be the canonical explosion mechanism of very massive stars.
From these, in the future study, one should address the issues like event rates of aspherical/spherical-like SN explosions as well as what kind of remnants are left behind, by considering various progenitor properties, e.g., mass, rotation, and magnetic fields, and compare with observations to eventually narrow down the MST parameters.
We also would like to mention that the present MST models might be constrained from their nucleosynthesis yields, possibly $r$-process elements, as they are similar to the QCD phase-transition driven SNe \cite{Fischer:2020}.
Exploring the parameter space, the progenitor model as well as the nucleosynthesis yields will be our future work.

\begin{acknowledgments}
We thank K. V. Aelst, D. Traykova, H.-J. Kuan, and A. T.-L. Lam for fruitful discussions.
This work was in part supported by Grant-in-Aid for Scientific Research 
(Nos. 20H00158 and 23H04900) of Japanese MEXT/JSPS.
Numerical computations were carried out on Sakura and Raven at Max Planck Computing and Data Facility and also on Cray XC50 at CfCA of NAOJ.
\end{acknowledgments}


\bibliographystyle{apsrev4-1}

\bibliography{apssamp}


\setcounter{equation}{0}
\setcounter{figure}{0}
\setcounter{table}{0}
\setcounter{page}{1}
\makeatletter
\renewcommand{\theequation}{S\arabic{equation}}
\renewcommand{\thefigure}{S\arabic{figure}}
\renewcommand{\bibnumfmt}[1]{[S#1]}
\renewcommand{\citenumfont}[1]{S#1}



\end{document}